\documentclass[aps,prl,twocolumn,showpacs]{revtex4}
\usepackage{epsfig}

\newcommand{\be}{\begin{equation}}
\newcommand{\ee}{\end{equation}}
\newcommand{\bc}{\begin{center}}
\newcommand{\ec}{\end{center}}
\newcommand{\bi}{\begin{itemize}}
\newcommand{\ei}{\end{itemize}}
\newcommand{\ba}{\begin{eqnarray}}
\newcommand{\ea}{\end{eqnarray}}

\newcommand{\ignore}[1]{}

\begin{document}

\title{Generic absorbing transition in coevolution dynamics.}

\author{Federico Vazquez}
\email[E-mail: ]{federico@ifisc.uib.es}
\author{V\'{\i}ctor M. Egu\'{\i}luz}
\author{Maxi San Miguel}

\affiliation{IFISC, Instituto de F\'{\i}sica Interdisciplinar y
Sistemas Complejos (CSIC-UIB), E-07122 Palma de Mallorca, Spain}

\date{\today}

\begin{abstract}

We study a coevolution voter model on a complex network. A mean-field
approximation reveals an absorbing transition from an active to a frozen phase
at a critical value $p_c=\frac{\mu-2}{\mu-1}$ that only depends on the average
degree $\mu$ of the network.  In finite-size systems, the active and frozen
phases correspond to a connected  and a fragmented network respectively.   The
transition can be seen as the sudden change in the trajectory  of an
equivalent random walk at the critical point, resulting in an approach to  the
final frozen state whose time scale diverges as $\tau \sim |p_c-p|^{-1}$  near
$p_c$.

\end{abstract}


\maketitle

The dynamics of collective phenomena in a system of interacting units depends
on both, the topology of the network of interactions and the interaction rule
among connected units. The effects of these two ingredients on the emergent
phenomena in a fixed network have been extensively studied. However, in many
instances, both the structure of the network and the dynamical processes on it
evolve in a coupled manner \cite{Zimmermann04,Gross07}.  In particular, in the
dynamics of social systems
(Refs.~\cite{Gross07,Kossinets06,Eguiluz05,Centola07} and references therein),
the network of interactions is not an exogenous structure, but it evolves and
adapts driven by the changes in the state of the nodes that form the
network. In recent models implementing this type of coevolution dynamics
\cite{Zimmermann04,Eguiluz05,Ehrhardt06,Benczik07,Gil06,Holme06,Vazquez07a,
Centola07,Nardini07,Kozma07} a transition is often observed from a phase where
all nodes are in the same state forming a single connected
network to a phase where the network is fragmented into disconnected
components, each composed by nodes in the same state \cite{Zachary77}.

In this paper we address the question of how generic is this type of
transition and which is the mechanism behind it. For this purpose, we
introduce a minimal model of interacting binary state nodes that incorporates
two basic features shared by many models displaying a fragmentation
transition: (i) two or more absorbing states in a fixed connected network, and
(ii) a rewiring rule that does not increase the number of links between nodes
in opposite state. The state dynamics consists of nodes copying the state of a
random neighbor, while the network dynamics results from nodes dropping their
links with opposite-state neighbors and creating new links with randomly
selected same-state nodes. This model can be thought as a coevolution version
of the voter model \cite{Holley75} in which agents may select their
interacting partners according to their states. It has the advantage of being
analytically tractable and allows a fundamental understanding of the network
fragmentation, explaining the transition numerically observed in related
models \cite{Gil06,Holme06,Vazquez07a,Centola07,Nardini07,Kozma07}. The
mechanism responsible for the transition is the competition between two
internal time scales, happening at a critical value that controls the relative
ratio of these scales.

\begin{figure}
\begin{center}
 \vspace*{0.cm}
 \includegraphics[width=0.45\textwidth]{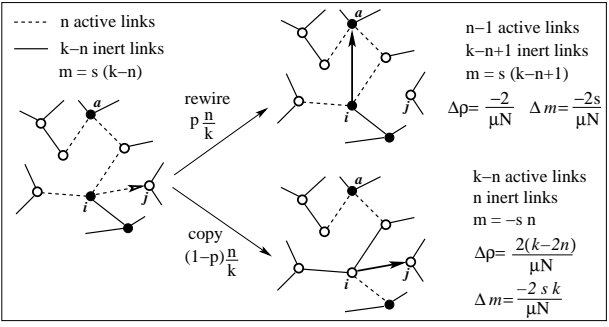}
 \caption{Update events and the associated changes in the density of active
 links $\rho$ and the link magnetization $m=\rho_{++} - \rho_{--}$ when two
 neighbors $i$ and $j$ with states $S_i=s$ and $S_j=-s$ are chosen
 ($s=\pm1$).}
 \label{voter}
\end{center}
\end{figure}

We consider a network with $N$ nodes. Initially, each node is endowed with a
state $+1$ or $-1$ with the same probability $1/2$, and it is randomly
connected to exactly $\mu$ neighbors, forming a network called degree-regular
random graph. In a single time step (see Fig.~\ref{voter}), a node $i$ with
state $S_i$ and one of its neighbors $j$ with state $S_j$ are chosen at
random, then:
\begin{enumerate}
 \item if $S_i = S_j$ nothing happens.
 \item if $S_i \not= S_j$, then with probability $p$, $i$ detaches its link to
     $j$ and attaches it to a randomly chosen node $a$ such that $S_a = S_i$
     and $a$ is not already connected to $i$; and with probability $1-p$, $i$
     adopts $j$'s state. 
\end{enumerate}
The \emph{rewiring probability} $p$ measures the rate at which the network
evolves compared to the rate at which the states of the nodes change; the
extreme values correspond to a fixed network ($p=0$), and to only rewiring
($p=1$).

\emph{Link Dynamics.} The evolution of the system can be described by the
densities of two different types of links: links connecting nodes with
different states or \emph{active links} and links between nodes in the same
state or \emph{inert links}. Note that an update (either rewire or copy) only
occurs when an active link is chosen.

\begin{figure}[t]
\begin{center}
 \vspace*{0.cm}
 \includegraphics[width=0.36\textwidth]{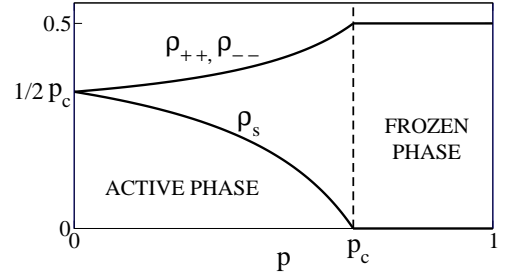}
 \caption{Stationary density of active links $\rho_s$ and the two types of
 inert links $\rho_{++}$ and $\rho_{--}$ vs the rewiring
 probability $p$ as described by the mean-field theory for a network
 with average degree $\mu=4$. The critical point $p_c$ separates
 an active from a frozen phase.}
 \label{rhos}
\end{center}
\end{figure}

In Fig.~\ref{voter} we describe the possible changes in the global density of
active links $\rho$ and their probabilities in a single time step, when a node
of degree $k$ is chosen. We denote by $n$ the number of active links connected
to node $i$ before the update. With probability $n/k$ an active link $i-j$ is
randomly selected. Then with probability $p$ the link $i-j$ is rewired and
becomes inert (link $i-a$), giving a local change of active links $\Delta
n=-1$ and a global density change of $\Delta \rho=-\frac{2}{\mu N}$, where
$\mu N /2$ is the total number of links, $\mu = \langle k \rangle = \sum_k k
P_k(t)$ is the number of links per node or average degree and $P_k(t)$ is the
node degree distribution at time $t$. (Even though $P_k(t)$ depends on time,
given that the network is constantly evolving, $\mu$ is constant because the
total number of links is conserved at each time step. Furthermore, simulations
show that $P_k(t)$ has a narrow shape with a maximum at $\mu$, as expected
from the random nature of the rewiring). With probability $(1-p)$ node $i$
flips its state changing the state of links around it from active to inert and
vice-versa, and leading to $\Delta n=k-2n$ and $\Delta \rho =
\frac{2(k-2n)}{\mu N}$. Assembling these factors, the change in the average
density of active links in a single time step of interval $dt=\frac{1}{N}$ is
described by the master equation
\begin{eqnarray}
\frac{d \rho}{dt}&=&\sum_k \frac{P_k}{1/N} \sum_{n=0}^k
B_{n,k}\,\frac{n}{k}\left[ (1-p) \frac{2 (k-2n)}{\mu N} - p \frac{2}{\mu N}
\right] \nonumber \\
\label{drkdt3}
&=& \sum_k P_k
\frac{2}{\mu k}\left[ (1-p) \left( k \langle n \rangle_k -2 \langle n^2
\rangle_k \right) - p \langle n \rangle_k \right],
\end{eqnarray}
where $B_{n,k}$ is the probability that $n$ active links are connected to a
node of degree $k$, and $\langle n \rangle_k$ and $\langle n^2 \rangle_k$ are
the first and the second moments of $B_{n,k}$ respectively.
In a mean-field (MF) spirit, in which the system is considered to be 
homogeneous, we approximate the probability that a given link is active by the 
average density $\rho$.  Then $B_{n,k}$ becomes the binomial distribution
with first and second moments $\langle n \rangle_k = \rho k$ and 
$\langle n^2 \rangle_k = \rho k + \rho^2 k(k-1)$.  Replacing 
these expressions in Eq.~(\ref{drkdt3}) we obtain a closed equation for the 
time evolution of $\rho$
\begin{equation}
\frac{d \rho}{dt}= \frac{2 \rho}{\mu} \left[ (1-p)(\mu-1)(1-2\rho)-1 \right].
\label{drdt}
\end{equation}
Equation~(\ref{drdt}) has two stationary solutions. For $p < p_c$,
the stable solution is
\begin{equation}
\rho_s \equiv \xi(p)=\frac{(1-p)(\mu-1)-1}{2(1-p)(\mu-1)},
\label{xi-p}
\end{equation}
corresponding to an active steady-state with a constant fraction of active
links in the system, while for $p>p_c$ the stable solution $\rho_s=0$
corresponds to an absorbing state where all links are inert. Thus, the MF
approach predicts an \emph{absorbing transition} (see Fig.~\ref{rhos}) from an
active to a frozen phase at a critical value
\begin{equation}
p_c=\frac{\mu-2}{\mu-1}.
\label{pc}
\end{equation}
A stationary state in the active phase, characterized by a density  $\xi(p) =
\frac {p_c-p}{2(1-p)}$ of active links, is composed by links continuously
being rewired (evolving network) and nodes flipping their states. For $p=0$
(original voter model), the value $\xi(0)=\frac{\mu-2}{2(\mu-1)}$ agrees very
well with the numerical values of the voter dynamics in different random
graphs \cite{Suchecki05a,Castellano05}. In the frozen phase, final states
correspond to a fixed network where connected nodes have the same state and no
more evolution is possible. The transition is continuous, with the order
parameter $\rho_s$ changing continuously at $p_c$. Close and below the
transition point, $\rho_s$ scales as $(p_c-p)$, thus the MF critical exponent
is $1$.

In order to obtain an insight about the structure of the network in both
phases we introduce $\rho_{++}$ ($\rho_{--}$) as the density of links
connecting two nodes with states $+1$ ($-1$).  It can be shown that
$\rho_{++}$ ($\rho_{--}$) is related to $\rho$ and the density $\sigma_+$
($\sigma_-$) of $+$ ($-$) nodes  by
\begin{equation}
\label{r++}
\rho_{++}=\sigma_{+}-\rho/2; \ \ \ \ \
\rho_{--}=\sigma_{-}-\rho/2.
\end{equation}
Due to the conservation of the ensemble average  of $\sigma_+$ and $\sigma_-$
for the voter model dynamics, we have that $\sigma_+=\sigma_-=1/2$, and
therefore, $\rho_{++}=\rho_{--}=\frac{1}{2}(1-\rho)$.  In the active phase the
continuous rewiring of links keeps the network connected in a single
component, i.e., a set of connected nodes. But, in the frozen phase only
inert links are present and in the same proportion
($\rho_{++}=\rho_{--}=1/2$), thus we expect the formation of two large
disconnected components with opposite state (see Fig.~\ref{rhos}). Therefore,
the MF description reveals a \emph{fragmentation transition} in the stationary
structure of the network, associated with the absorbing transition at $p_c$.

\begin{figure}[t]
\begin{center}
 \vspace*{0.cm}
 \includegraphics[width=0.48\textwidth]{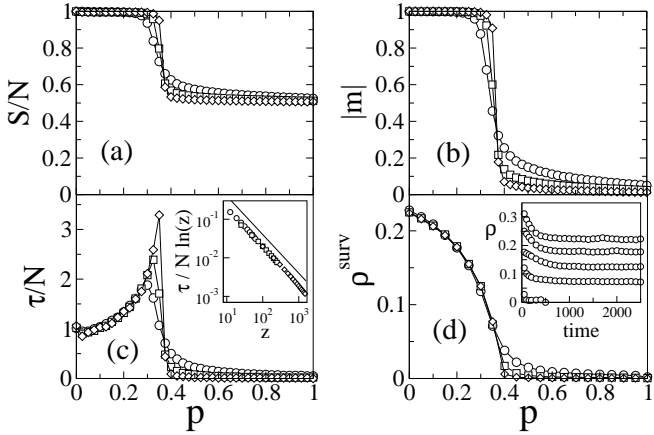}
 \caption{(a) Average relative size of the largest network component $S$ and
 (b) absolute value of the link magnetization $m$ vs $p$ in the final frozen
 state. (c) Average convergence time $\tau$ per system size $N$ vs $p$.
 Inset: scaling of $\tau$ for $p \gtrsim p_c \simeq 0.38$, indicating that
 $\tau \sim \frac{N}{z} \ln(z)$ with $z=\mu (p-p_c)N$. The
 solid line has
 slope $-1$. (d) Stationary density of active links in surviving runs
 $\rho^{\mbox{\scriptsize surv}}$. Inset: average time evolution of
 $\rho$. The averages are over $10^4$ realizations of networks with
 $\mu=4$ and sizes $N=250$ (circles), $1000$ (squares) and $4000$
 (diamonds).}
 \label{T-m-S}
\end{center}
\end{figure}

\emph{Final states in a finite system.} The previous MF approach predicts a
transition in the limit of an infinite large network.  However, for any value
of $p$, due to fluctuations, a finite size network eventually reaches an
absorbing state composed by inert links only. We studied the structure of the
network in the final state by performing numerical simulations of the dynamics
starting with a degree-regular random graph with connectivity $\mu=4$ and
letting the system evolve until it was frozen. In Fig.~\ref{T-m-S}(a) we plot
the average size of the largest network component $S$ in the final
configuration for networks with $N=250,1000$ and $4000$ nodes. We observe that
$S$ is very close to $N$ for values of $p$ below a transition point $p_c
\simeq 0.38$, indicating that the network forms a single component
\cite{pc}. Above $p_c$ the network gets disconnected into two large components
and a set of components of size much smaller than $N$, giving a value of $S
\simeq N/2$.

To compare simulations with MF results, we calculated the stationary density
of active links in surviving runs $\rho^{\mbox{\scriptsize surv}}$. As we show
in Fig.~\ref{T-m-S}(d), $\rho^{\mbox{\scriptsize surv}}$ monotonically
decreases with $p$, becoming sharper with increasing system and indicating a
transition from an active to a frozen phase as predicted by the MF theory. In
the active phase, $\rho^{\mbox{\scriptsize surv}}$ reaches a steady value
larger than zero and independent on the system size $N$, while in the frozen
phase $\rho^{\mbox{\scriptsize surv}}$ vanishes in the thermodynamic
limit. The critical point for the active-frozen transition $p_c \simeq 0.38$
calculated from Fig.~\ref{T-m-S}(d) is roughly the same as for the
fragmentation transition (Fig.~\ref{T-m-S}(a)), suggesting that the active and
frozen phases observed in infinite large systems correspond to the connected
and disconnected phases respectively in finite systems.  The MF critical point
$p_c=2/3$ calculated using Eq.~(\ref{pc}) with $\mu=4$ differs from the
numerical value $p_c \simeq 0.38$ (Figs.~\ref{T-m-S}(a,d)) due to correlations
appearing in the rewiring process. These correlations, that are not taken into
account in the analytical approximation, make the moments $\langle n \rangle_k$
and $\langle n^2 \rangle_k$ different from the analytical
values $\rho k$ and $\rho k + \rho^2 k(k-1)$ respectively. These deviations
have the overall effect of decreasing the observed critical point respect to
the theoretical one.

\emph{Approach to the absorbing states.} So far, we have shown that a finite
network under the coevolving dynamics experiments a fragmentation transition
as the rewiring rate is increased. We now unveil the mechanism of this 
transition by studying the evolution of the system to the frozen state.

We represent the state of the system as a point $(m,\rho)$ in the 2
dimensional space, where the coordinates are the \emph{link magnetization} $m
= \rho_{++}-\rho_{--}$ and the density of active links respectively. When a
node of degree $k$ connected to $n$ active links is chosen, the possible
changes in $m$ and $\rho$ and their respective probabilities are those
described in Fig.~\ref{voter}. In the $(m,\rho$) space, the system undergoes a
random walk (RW) inside the triangle $0 \le \rho + |m| \le 1$, whose
boundaries follow from the constraint relation $\rho_{--} + \rho_{++} + \rho =
1$. The system reaches an absorbing configuration and stops evolving when the
RW hits either one of the fixed points $(-1,0)$ or $(1,0)$ (all nodes in state
$-$ or $+$ respectively) or a point on the fixed line $\rho=0$ (frozen mixture
of $-$ and $+$ nodes). At the point $(-1,0)$ ($(1,0)$) only $--$ ($++$) links
are present, the network is composed by a giant component, and the system is
in the connected phase. Points on the line $\rho=0$ and close to point $(0,0)$
correspond to a frozen network with similar number of $++$ and $--$ links
arranged in two large $+$ and $-$ components (disconnected phase).

\begin{figure}[t]
\begin{center}
 \vspace*{0.cm}
 \includegraphics[width=0.44\textwidth]{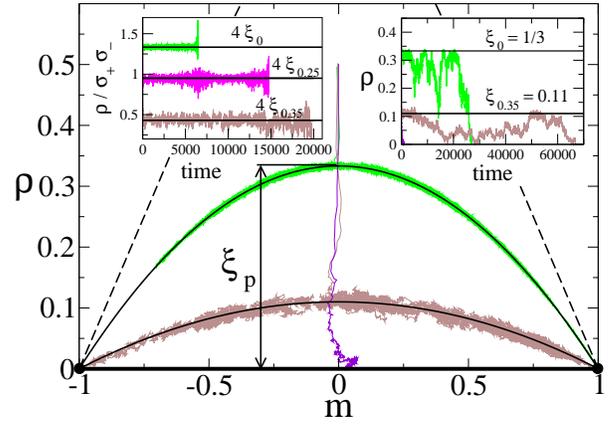}
 \caption{Typical trajectories of the random walk for a network of size
 $N=10^4$ and $\mu=4$. The upper ($p=0$) and lower ($p=0.35$)
 parabolas are the trajectories for rewiring rates below the transition
 point $p \simeq 0.38$, while the quasi-vertical line is for $p=0.4$.
 Insets: Time evolution of the density of active links $\rho$ (right) and
 the ratio between $\rho$ and the product $\sigma_+ \sigma_-$ (left) in a 
single realization for different values of $p$.}
 \label{r-m}
\end{center}
\end{figure}

In Fig.~\ref{r-m} we plot trajectories of the RW in one realization for
different values of $p$. For $p<p_c \simeq 0.38$, the motion of the RW has two
stages. In the first and very short stage, the RW travels along the $m \simeq
0$ axis from the starting point $(m \simeq 0, \rho \simeq 1/2)$ to the point
$(m \simeq 0, \rho \simeq \xi_p)$ that corresponds to the steady state $\rho_s
= \xi_p$ in infinite large systems (see right inset of Fig.\ref{r-m}).  In the
second and long stage, the RW diffuses on the $m$ direction, corresponding to
the fluctuations of $\rho$ down the steady state $\rho_s = \xi_p$, until
it hits either point $m=-1$ or $m=1$ (see Fig.~\ref{T-m-S}(b)). We observe in
Fig.~\ref{r-m} that the motion of the RW is not completely random but its
trajectory fluctuates around a curve described by
\begin{equation}
\rho_p(m) = \xi_p (1-m^2).
\label{rm}
\end{equation}
The origin of this relation is that both $\rho$ and $m$ can be expressed as
functions of $\sigma_+$.  As we observe in the left inset of  Fig.~\ref{r-m},
during one realization the ratio $\frac{\rho}{\sigma_+ \sigma_-}$ fluctuates
around the constant value $4\,\xi_p$.  Then, using Eq.~(\ref{r++}) we obtain
$\rho = 4 \,\xi_p \, \sigma_+ (1-\sigma_+)$ and $m = \rho_{++}-\rho_{--} =
2\sigma_+ -1$, from where we arrive to Eq.~(\ref{rm}) by eliminating
$\sigma_+$. For $p>p_c$, the bias to the $\rho=0$ line makes the RW hit a
point close to the origin (see the $p=0.4$ trajectory in
Fig.~\ref{r-m}). Simulations show that for a fixed value $p<p_c$, the
amplitude of the fluctuations of the RW's trajectory around its mean value
$\rho_p(m)$ vanishes as $N$ increases. Thus, increasing $N$ has the effect of
increasing the probability that the RW reaches one of the end points $m=\pm 1$
before it hits the line $\rho=0$, and therefore, most realizations end in a
single component.  Eventually, in the large $N$ limit, the RW has three
absorbing points: either point $(-1,0)$ or $(1,0)$ (single component network)
when $p<p_c$, and point $(0,0)$ (two components network) when $p>p_c$.

\emph{Convergence times.} A magnitude of interest is the average
time $\tau$ to reach an absorbing state. For $p<p_c$, the
$m$-coordinate of the walker performs a 1d symmetric random walk
with an average jumping interval and its probability that scale as
$1/N$ and $\rho \sim \xi$ respectively. To reach one of the end
points $m=\pm 1$ the RW needs to attempt an average of $N^2/\xi$ steps, 
and given that the time increases by $1/N$ in each attempt,
we find that $\tau \sim N/\xi$. From Eq.~(\ref{drdt}) for $p
\gtrsim p_c$, $\rho$ decays as $\rho(t) \sim - \xi \,
e^{\,4 \, \xi\, t/\mu}$. The system freezes at a time $\tau$ for
which $\rho(\tau) \sim 1/\mu\,N$, then 
$\tau \sim -\frac{\mu}{4\xi} \ln(-\mu \, \xi \, N)$.
Using the MF
approximation $\xi(p) \sim (p_c-p)$ close to $p_c$, we obtain that
$\tau \sim N(p_c-p)^{-1}$ as $p \to p_c^-$ and $\tau \sim
(p-p_c)^{-1}\ln [\mu (p-p_c)N]$ as $p \to p_c^{+}$, thus the
convergence to the final state slows down at the critical point
(see Fig.\ref{T-m-S}(c)).

{\em Summary and conclusions.} In summary, the coevolution
mechanism on the voter model induces a fragmentation transition that is a
consequence of the competition between the copying and the
rewiring dynamics. In the connected active phase, the system falls
in a dynamical steady state with a finite fraction of active
links. The slow and permanent rewiring of these links keeps the
network evolving and connected until by a finite-size fluctuation
the system reaches the fully ordered state (all nodes in the same
state) and freezes in a single component. In the frozen phase, the
fast rewiring dynamics quickly leads to the fragmentation of the
network into two components, before the system becomes fully
ordered.

The similarity between the mean-field equation for the density of
active links in the coevolution voter model (Eq.~\ref{drdt}) and
the one for the density of infected sites in the contact process
\cite{Marro99}, suggests that our model could belong to the
Directed Percolation universality class. However, both models seem
not to be equivalent given that our model possesses many absorbing
states (any point on the $\rho=0$ line of Fig.~\ref{r-m}), unlike
the contact process where there is a single absorbing state
characterized by the absence of infected sites. We believe that our results
provide a new insight in the ongoing discussion about models with
infinitely many absorbing states \cite{Park07}.

We thank E. V. Albano for useful comments on dynamical critical
behavior,  and financial
support from MEC (Spain), CSIC (Spain) and EU  through projects 
FISICOS, PIE200750I016, and GABA, respectively.


{}



\begin{thebibliography}{}
\bibitem{pc} We have also calculated the critical point by means
of the survival probability \cite{Munoz99}. Setting as initial
condition a configuration with a few active links, the survival
probability at time $t$ is calculated as the fraction of
realizations that have not reached an absorbing state by time $t$.
At the critical point ($p_c \simeq 0.38$), the survival
probability displays a power law distribution.
\end{thebibliography}
\end{document}